\documentclass[modern]{aastex62}

\usepackage{color}

\shorttitle{Searching for hypermassive neutron stars}

\begin{document}

\title{Searching for hypermassive neutron stars with short gamma-ray bursts}

\correspondingauthor{Cecilia Chirenti}
\email{cecilia.chirenti@ufabc.edu.br}

\author{Cecilia Chirenti}
\affil{Centro de Matem\'atica, Computa\c c\~ao e Cogni\c c\~ao, UFABC, 09210-170 Santo Andr\'e-SP, Brazil}

\author{M. Coleman Miller}
\affiliation{Department of Astronomy, University of Maryland, College Park, MD 20742-2421, USA}
\affiliation{Joint Space-Science Institute, University of Maryland, College Park, MD 20742-2421, USA}

\author{Tod Strohmayer}
\affiliation{Joint Space-Science Institute, University of Maryland, College Park, MD 20742-2421, USA}
\affiliation{Astrophysics Science Division and Joint Space-Science Institute\\
NASAs Goddard Space Flight Center, Greenbelt, MD 20771, USA}

\author{Jordan Camp}
\affiliation{Astrophysics Science Division and Joint Space-Science Institute\\
NASAs Goddard Space Flight Center, Greenbelt, MD 20771, USA}

\begin{abstract}

Neutron star mergers can form a hypermassive neutron star (HMNS) remnant, which may be the engine of a short gamma ray burst (SGRB) before it collapses to a black hole, possibly several  hundred milliseconds after the merger. During the lifetime of a HMNS, numerical relativity simulations indicate that it will undergo strong oscillations and emit gravitational waves with frequencies of a few kilohertz, which are unfortunately too high for detection to be probable with Advanced LIGO. Here we discuss the current and future prospects for detecting these frequencies as modulation of the SGRB. The understanding of the physical mechanism responsible for the HMNS oscillations will provide information on the equation of state of the hot HMNS, and the observation of these frequencies in the SGRB data would give us insight into the emission mechanism of the SGRB.

\end{abstract}


\section{Introduction} \label{sec:intro}

The observation of the first binary neutron star merger GW170817 using LIGO and Virgo \citep{2017PhRvL.119p1101A} and the associated short gamma-ray burst (SGRB) GRB170817A \citep{2017ApJ...848L..13A} brought a wealth of information not only in gravitational waves (GWs) but also in the electromagnetic counterpart of the signal 
(see \citealt{2017ApJ...848L..12A} for an early summary), which led the scientific community into the era of multimessenger astronomy with GWs. Although this ``golden binary'' observation was an extremely fortunate event, we have every reason to be even more optimistic for O3 (the current LIGO/Virgo run that started on April 1st) and for the future. The detection sensitivity has been increased, and in the first  months there has been roughly one detection per week of a compact binary coalescence.  Furthermore the Japanese detector KAGRA is expected to join toward the end of the run, which will result in even better data and sky localization. 

The merger of two neutron stars has long been proposed as one of the possible progenitors of SGRBs 
(see \citealt{2014ARA&A..52...43B} for a recent review).  Depending on the combined masses of the neutron stars and on the maximum mass of a neutron star, there are in principle four possible outcomes to the merger:

\begin{enumerate}

\item Prompt formation of a black hole.  In this scenario, the total mass is too large to be sustained by rotation of any type and thus a black hole forms on essentially a free-fall time.  If the two neutron stars had nearly equal masses then tidal tails will contain little mass and thus the matter that remains outside the horizon will likely be insufficient to drive a SGRB (see \citealt{2008PhRvD..78h4033B} for a discussion of this point).  However, if there exist higher-mass versions of the asymmetric double neutron star binary PSR~J1453+1559 (with an estimated pulsar mass of $1.559\pm 0.005~M_\odot$ and companion mass of $1.174\pm 0.004~M_\odot$; see \citealt{2015ApJ...812..143M}) then potentially there could be sufficient material outside the black hole to power a SGRB.

\item Formation of a hypermassive neutron star (HMNS), which is defined as a star that is temporarily supported against collapse by strong differential rotation but that is above the maximum mass that can by supported by uniform rotation \citep{2000ApJ...528L..29B}. It is expected that within tens to hundreds of milliseconds after the merger the star will lose angular momentum due to the emission of GWs, finally collapsing to a black hole \citep{2000ApJ...544..397S}. HMNSs and their surrounding accretion disks are strong candidates for the engines of SGRBs (e.g., \citealt{2006PhRvL..96c1102S} and \citealt{2008PhRvD..78h4033B}).

\item Formation of a supramassive neutron star, which is defined as a star that can be held up against collapse by uniform rotation but that is above the maximum mass for a non-rotating star.  Such a star remains stable as long as its angular momentum is sufficient to prevent collapse, and thus can last for seconds to years.  In this case, it is expected that the merger will produce a rapidly rotating, highly magnetized neutron star (i.e., a millisecond magnetar) that can inject energy into the burst \citep{2008MNRAS.385.1455M}.

\item Formation of a stable neutron star.  The recent determination that PSR~J0740+6620 has a mass of $2.17^{+0.11}_{-0.10}~M_\odot$ \citep{2019arXiv190406759T}, combined with the existence of low-mass neutron stars such as the $1.174\pm 0.004~M_\odot$ mass of the companion to PSR~J1453+1559, suggests that in possibly rare circumstances the combined mass of the two neutron stars could be less than the maximum mass of a slowly rotating star.  

\end{enumerate}

\citet{2014ApJ...788L...8M} argue that if the remnant lasts more than one hundred milliseconds, the production of a wind due to neutrino emission will produce either a choked jet or a much longer-lasting gamma-ray event than is seen in SGRBs.  Thus they argue that 
SGRBs likely involve rapid ($<0.1$~seconds) collapse to a black hole, with 
a possible HMNS phase.\footnote{\citet{2014ApJ...788L...8M} also leave open the possibility 
of magnetar-driven SGRBs.  In that case, they require that the jet be 
launched $<0.1$~seconds after the collapse to avoid choking.} Post-merger observations of GW170817 also seem to support a HMNS phase (e.g., \citealt{2017PhRvD..96l3012S}, \citealt{2018ApJ...852L..25R}, \citealt{2018PhRvD..97b1501R}, and \citealt{2018ApJ...852L..29R}). For example, \cite{2014MNRAS.441.3444M} propose that early optical emission days after the merger is a sign of delayed black hole formation: the higher abundance of neutrinos generated in the merger (as compared with the case of a prompt black hole formation and appearance of an event horizon) raises the electron fraction and reduces the formation of lanthanides. The resulting material is rich in elements from the iron group, which have comparatively low opacity and are thought to be responsible for the ``early blue bump" seen within the first few days after GW170817 \citep{2017ApJ...848L..12A}.

Numerical relativity simulations also show that the HMNS should emit strongly in GWs, with a few 1-4 kHz peaks in the signal (see for instance \citealt{2014PhRvD..90b3002B,2014PhRvL.113i1104T}), whose physical origin is not yet completely understood. The detection of these frequencies would provide strong evidence for the HMNS phase and consequently information about the equation of state (EOS) in a hot and magnetized state that will not be probed by studies of GWs from the inspiral \citep{2018PhRvL.121p1101A}. Unfortunately they are in a frequency range too high (1-4 kHz) for realistic prospects of detection with current GW detectors, but they will be easily seen in the future with third generation GW  detectors such as the Einstein Telescope (ET; \citealt{2010CQGra..27s4002P}) and the Cosmic Explorer (CE; \citealt{2017CQGra..34d4001A}), which are expected to go online in approximately 15 years. 

However, we may not have to wait for third generation detectors. This signature of a HMNS phase may already be detectable in the electromagnetic counterpart of the signal, as a modulation of the SGRB. This hypothesis can be tested with existing SGRB data from the gamma-ray monitors BATSE \citep{2000ApJS..126...19P}, Fermi GBM \citep{2009ApJ...702..791M}, and Swift BAT \citep{2005SSRv..120..143B}.  At the same time, it is important to determine the prospects for detectability with proposed missions such as TAP \citep{TAPConceptStudyReport} and STROBE-X \citep{2019arXiv190303035R}. Moreover, a detection of the HMNS frequencies in the electromagnetic spectrum in coincidence with a GW detection of a binary neutron star merger could be used to guide a search for the frequencies in the GW signal with a lowered threshold, perhaps allowing their detection with advanced LIGO.

In this letter we will discuss this observational scenario in Section \ref{sec:modulation} and present some order of magnitude estimates for the detectability and the statistical significance of the expected SGRB modulation in Section \ref{sec:estimate}. We present our final remarks in Section \ref{sec:conclusions}.

\section{Modulation of the SGRB} \label{sec:modulation}

\begin{figure}[htbp] 
   \centering
   \includegraphics[width=2.05in, angle=-90]{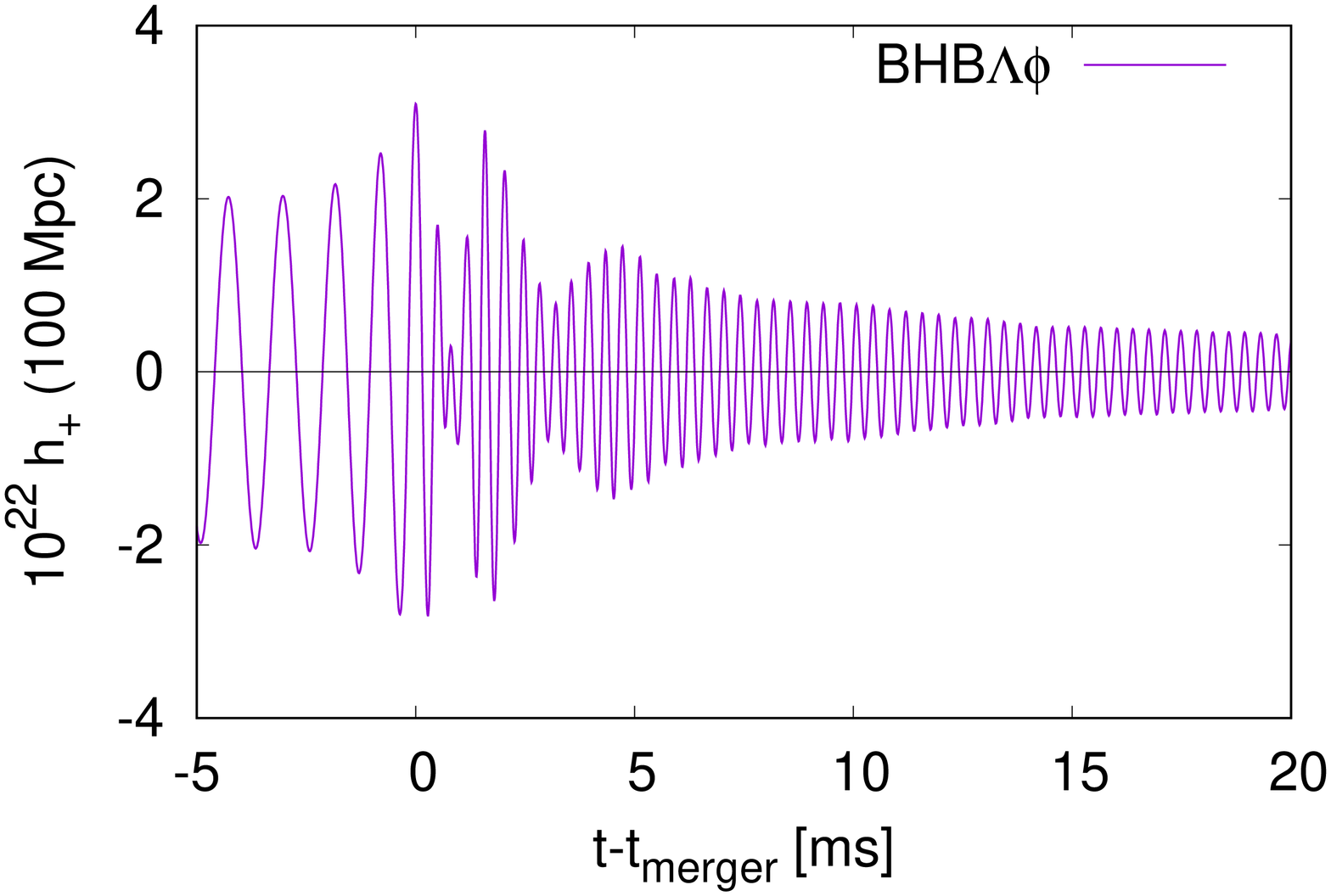} 
   \includegraphics[width=2.05in, angle=-90]{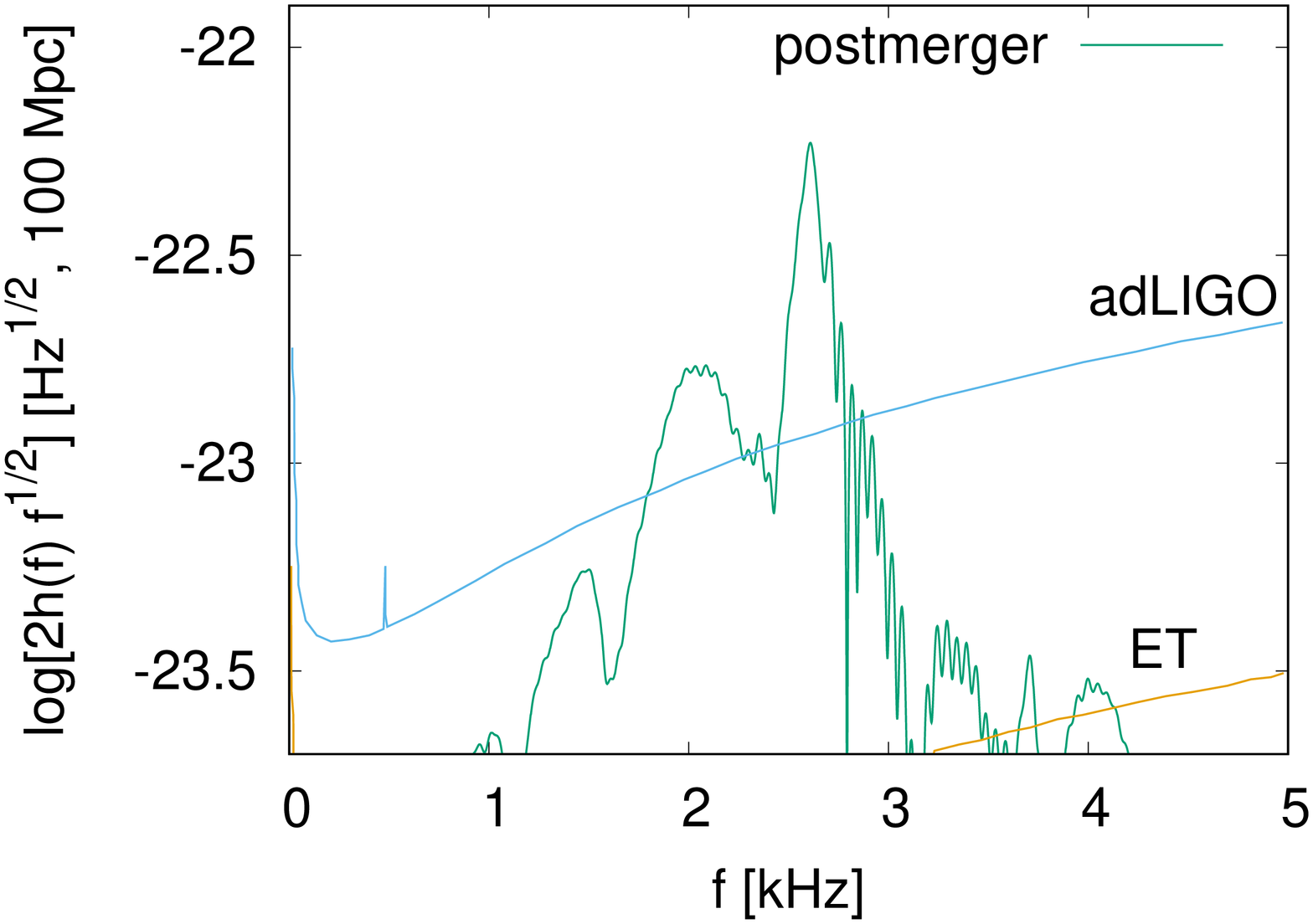} 
   \caption{Example of a postmerger GW signal of a long lived HMNS (left) and of its frequency spectrum (right), together with the predicted sensitivity curves for Advanced LIGO \citep{2010CQGra..27h4006H} and the Einstein Telescope \citep{2010CQGra..27a5003H}, both with SNR = 1.  The data are from the $1.35+1.35~M_\odot$ simulation of \citet{2017ApJ...842L..10R} (top left panel of their Figure~2), and were kindly provided by David Radice; many other simulations get similar results, e.g., \citet{2012PhRvL.108a1101B} and \citet{2014PhRvL.113i1104T}. The spectrum shows a couple of clear peaks in a complex structure; at least some of these peaks may be related to oscillations of the HMNS.  
   }
   \label{fig1}
\end{figure}

In Figure~\ref{fig1} we display a typical GW signal from a NS-NS merger resulting in a long-lived HMNS. The spectrum can show several complicated features, with at least a couple of clear peaks. The physical interpretation of different features in the spectrum is still not clear, although different correlations have been found, for instance relating the values of the frequencies of the main peak with the radius of the corresponding 1.6 $M_{\odot}$ star \citep{2012PhRvL.108a1101B}, to the tidal coupling constant \citep{2015PhRvL.115i1101B} and to the maximum instantaneous angular frequency of the differentially rotating HMNS \citep{2015PhRvD..91f4027K}, among other findings.

Given the well-established theory of stellar oscillations (see \citealt{1999LRR.....2....2K} for a review), and the general features of the GW frequencies observed in the simulations, it is possible that some of the peaks shown in Figure \ref{fig1} represent characteristic modes of oscillation of the HMNS, consistent for instance with the analysis performed by \cite{2011MNRAS.418..427S}, who have shown that the main peak is due to the $m=2$ f-mode (see also \citealt{2016EPJA...52...56B} and other works). 

The emission of a SGRB by a HMNS could in principle carry information from the strong oscillations of the star in this phase. \citet{1992PhDT........23S} presented an argument based on relativistic beaming to estimate the surface oscillation amplitude required to produce potentially observable variations in the beaming angle of radio pulsar emission. In more detail, the connection between the neutron star oscillations and the modulation of the beaming angle of the emission is realized by the strong magnetic field, as the ``shaking" of the magnetic field at the surface of the star by surface displacements caused by the stellar oscillations might perturb or modulate the emitting region above the stellar surface (see also \citealt{1976ApJ...208L..43B}). Therefore the effect of HMNS oscillations could result in a measurable modulation of the SGRB even though the jet has to make its way through the ejecta (see \citealt{1996ApJ...473..998F} for a study of long complex bursts which can be attributed to the central engine variability).

Therefore we expect that the high-frequency oscillations of the HMNS in the approximate range $1 - 4$ kHz could be observable, \emph{if the SGRB is emitted during the HMNS phase}. So far, most simulations including magnetic fields focused on jet formation from black hole remnants. Indeed, \cite{2016ApJ...824L...6R} find a mildly relativistic outflow (an incipient jet) only after a black hole is formed. The alternative possibility of a (hypermassive) magnetar central engine for the SGRB after a binary neutron star merger has been relatively little explored, and the relevant timescales for jet formation can be significantly longer than in the black hole + disk case. Current numerical relativity simulations have a common resolution limitation that is not sufficient to fully account for the main magnetic field amplification mechanisms \citep{2019PhRvD.100b3005C}.

The search for a high frequency modulation in a SGRB associated with a binary neutron star merger event could also serve as a probe for the emission mechanism of the burst. The absence of these frequencies in the data could point to the emission of the SGRB as the HMNS collapses to a black hole, which is one possible explanation for the observed $\approx 2$ s delay between GW170817 and GRB170817A \citep{2041-8205-848-2-L13}, but several other possibilities for this delay have been proposed in the literature, such as the time needed for the jet to reach the photosphere and hence be able to emit the SGRB.

Instruments suitable for the observation of electromagnetic transients with high timing resolution, such as the currently operating Fermi and Swift, and concept studies such as TAP and STROBE-X, could detect the modulation in the SGRB caused by the HMNS oscillations nearly in coincidence with future GW detections, enabled by continued increases in LIGO sensitivity.

Additionally, as we argue in the next section, signatures of HMNS oscillations might be present in extant data from especially bright and close SGRBs recorded by BATSE, Fermi and Swift.  A limited number of studies have searched for periodicity in gamma-ray emission.  For example, \citet{2002ApJ...576..932K} estimated that a 10 \% modulation amplitude would be detected $1/2$ of the time with their procedure. However, they found no evidence of periodic modulation in the $400-2500$~Hz range from BATSE data on more than 2000 gamma-ray bursts and more than 150 soft gamma-ray repeater flares. \citet{2013ApJ...777..132D} also had no detections in their $10-30$~Hz analysis of 44 bright SGRBs, but the expected frequencies of HMNS oscillations can be greater than the range that has been searched. More recently, \cite{2018ApJ...855..101H} looked at the TTE BATSE data (time-tagged event), but they were mostly interested in the structure (shape) of pulses of
emission and restricted the resolution to 4 ms.

Perhaps the most important point to notice is that the rapid evolution of the differential rotation inside an HMNS will cause the characteristic frequencies to evolve during the burst and therefore strict periodicity is not expected.  Thus a search will require a careful analysis of the expected frequency evolution, which we defer to a later treatment.  In the next section we give a broad motivation for why such oscillations are in principle detectable.

\section{Detectability and statistical significance of the modulation}
\label{sec:estimate}

Taking as a representative example the dominant frequency peak at $f \approx 2.5\ {\rm kHz}$ from the right panel of Fig. \ref{fig1}, we can use an approximate expression for the GW strain amplitude from a pulsar of period $P$ at a distance $r$ to calculate the associated surface displacement $\Delta R$ needed for an oscillation mode to produce those signals. The GW strain amplitude in this approximation is given by
\begin{equation}
h \approx 4 \times 10^{-23}\epsilon \ (P/1\, {\rm ms})^{-2}(100\, {\rm Mpc}/r)\,,
\label{eq:h}
\end{equation}
where we are modelling the star as an ellipsoid with semi-major axes {\bf $a>b>c$} rotating around its minor axis and $\epsilon$ is the ellipticity in the equatorial plane, 
 defined as $\epsilon = (a-b)/(ab)^{1/2}$. From this simple model, taking $P = 2/f$ and using the simulation data we find $\epsilon \approx 8.5 \times 10^{-3}$ and $\Delta R \equiv a - b \approx \frac{\sqrt{2}}{2}\epsilon R \approx 120\ {\rm m}$, assuming a representative HMNS radius of approximately 20 km \citep{2017PhRvD..95f3016C}.

Motivated by the analysis of \cite{1992PhDT........23S} we can propose that, for any arbitrary oscillation mode, the maximum variation possible for the deviation $\Delta \theta$ of the SGRB beam direction will be roughly the slope of the perturbation at the surface, given by the surface displacement $\Delta R$ and the wavelength $\lambda$ of the mode as $\Delta \theta \approx \Delta R/(\lambda/4)$. 
If the sound speed is the 
$c/\sqrt{3}$ characteristic of high energy density matter then 
$\lambda\approx c/\sqrt{3}f$, and thus $\Delta\theta\approx 7\times 
10^{-3}$.

This deviation must be compared with the larger between the relativistic beaming angle and the jet opening angle of the SGRB. The relativistic beaming angle is $\theta_b \approx 1/\gamma$, where $\gamma$ is the relativistic Lorentz factor of the flow, and for typical cases of GRBs we have $\gamma \approx 10^2-10^3$, and therefore $\theta_b \approx 10^{-3}-10^{-2}$. Typical values for the jet half-opening angle are $\theta_j \approx 0.1$, but lower values of $\theta_j \approx 0.02$ have also been reported \citep{Jin_2018}. As a result, we expect that in the population of SGRBs, HMNS oscillations can produce $\Delta \theta$ up to $ \approx 0.4\, \theta_j$. If we estimate that the flux variation should be $\approx \Delta \theta/\theta_j$,  this mechanism should produce a noticeable modulation of the signal with tens of percent of flux variation.

We estimate the number $n$ of SGRB photon counts during the lifetime of a HMNS 
\begin{equation}
n = F_{\rm SGRB} \times \Delta T_{\rm HMNS} \times A_{\rm det}/E_{\rm peak}^{\rm obs}\,,
\label{eq:n}
\end{equation}
where we use average values for SGRBs: $F_{\rm SGRB} \approx 5 \times 10^{-6}\ {\rm erg}\ {\rm cm}^{-2}\ {\rm s}^{-1}$ (flux of a moderately bright burst) and $E_{\rm peak}^{\rm obs} \approx 350\ {\rm keV}$ (observed energy at the peak; \citealt{2009A&A...496..585G}) and $\Delta T_{\rm HMNS} \approx 0.1\ {\rm s}$ (lifetime of a HMNS;  \citealt{2014PhRvD..90b3002B,2014PhRvL.113i1104T}). It is worth mentioning here that our approach would be unchanged in the case of a supramassive neutron star instead of a HMNS; the only difference would be a longer lifetime leading to an even more optimistic estimate. Using values for the effective detector area, $n$ is 
approximately 1780 for BATSE \citep{2000ApJS..126...19P}, 1250 for Swift \citep{2005SSRv..120..143B}, 110 for Fermi \citep{2009ApJ...702..791M} and $n \approx 790$ for proposed mission TAP \citep{TAPConceptStudyReport}. STROBE-X will have a much larger area than the other detectors (about 4 $\rm{m}^2$) but will be limited to lower energies \citep{2019arXiv190303035R}; however, the Band model for GRB spectra \citep{1993ApJ...413..281B} has a low-energy spectral index $\alpha$ close to -0.4 for SGRBs (see for instance \citealt{2009A&A...496..585G}). If we extrapolate this spectrum to 10 keV from the 30 keV lower limit of the BATSE Large Area Detector (LAD) data used by \cite{2009A&A...496..585G}, we would find typically ${n\approx4230}$ counts in STROBE-X.\footnote{Here we have assumed that the burst was detected directly by STROBE-X. However, the large area field of view of the LAD is small ($\approx$ 1 deg collimated) and a direct detection would be unlikely. A more likely scenario would be a burst outside the field of view, in which case only a fraction of the photons would reach the detector, with an unknown reduction factor in the effective area of the detector.}

The expected statistical fluctuation in the photon count is $\sqrt{n}$, which gives a relative fluctuation of approximately $2\%-10\%$. This is significantly lower than the relative fluctuation of up to $\sim 50$\% that we expect to be caused by the modulation of the signal due to the HMNS oscillation. Consequently, even if the efficiency of the mechanism we propose results in a significantly smaller relative modulation, it is potentially observable. 

An apparent concern would be that, given the expected frequencies of a few kHz, there would not be enough photon counts in the small time bins needed to resolve the period of a HMNS oscillation. However, as \cite{1988SSRv...46..273L} point out, when the background is weak compared with the source the confidence level in terms of sigmas at which a feature corresponding to a signal with a fractional variation $a_{\rm osc}$ (due to an oscillation) will be detected can be estimated by
\begin{equation}
n_{\sigma} = \frac{1}{2}Ia^2_{\rm osc}\sqrt{\frac{\Delta T}{\Delta f}}\,,
\label{eq:n}
\end{equation}
where $I$ is the source count rate, $\Delta T$ is the total observing time and $\Delta f$ is the frequency width of the peak in the Fourier spectrum. \footnote{In cases when the background is comparable to the signal, the source count rate $I$ in eq. (\ref{eq:n}) must be replaced with $I^2/(I + B)$, where $B$ is the background count rate. However, it is common though not universal to have $I \gg B$ in SGRBs; see \cite{2018ApJ...855..101H} for example light curves.} Therefore SGRB data 
can be searched for the HMNS oscillations even if the number of counts per time resolution element is small. 
Using the values estimated with eq. (\ref{eq:n}), we find that an oscillation with a fractional variation $a_{\rm osc} = 0.25$ would be detectable at the $11 \sigma$ level by BATSE and at $8 \sigma$ by Swift.  The proposed missions TAP and STROBE-X will be able to detect the signal at the $5 \sigma$ and $26 \sigma$ level, respectively. Oscillations in an event with a flux three times higher than the average estimate of \cite{2009A&A...496..585G} (compatible with GRB 120323A)
would be detectable by Fermi at over $5 \sigma$  with a stronger fractional variation of $a_{\rm osc} = 0.4$. 

\section{Final remarks}
\label{sec:conclusions}

We have presented a preliminary analysis of the detectability of HMNS oscillations as modulation of SGRB signal emitted in the electromagnetic counterpart of a binary neutron star merger, showing promising results. Archival data from gamma-ray detectors can be searched for these signals, as well as future data obtained in coincidence with GW detections. However, the analysis of existing and future data should be performed carefully, as the frequencies may drift as the HMNS spins down during its lifetime.

Our analysis assumes that the SGRB is emitted during the HMNS phase after the merger. Therefore the presence of these frequencies in the signal will favor the HMNS scenario for SGRB emission, whereas their absence would support scenarios involving prompt collapse. 

The detection of frequencies corresponding to HMNS oscillations will provide information about the hot EOS after the merger, which cannot be probed by tidal deformability effects on the GW signal during the inspiral (prior to the merger). Additionally, if a SGRB is detected in coincidence with a future GW detection, it could facilitate a GW search for the HMNS oscillations with a lower detection threshold.  

\acknowledgments

The authors thank Simone Dichiara, Bernard Kelly, Amy Lien, Scott Noble, Luciano Rezzolla, Bernard Schutz and Jeremy Schnittman for useful discussions and comments on a previous version of this manuscript, and David Radice for kindly providing the NR simulation data.  MCM was at the Kavli Institute for Theoretical Physics during the completion of this paper and was therefore supported in part by the National Science Foundation under Grant No. NSF PHY-1748958. This work was supported in part by the Munich Institute for Astro- and Particle Physics (MIAPP) of the DFG cluster of excellence ``Origin and Structure of the Universe" and by the Brazilian National Council for Scientific and Technological Development (CNPq grant 303750/2017-0). 

\bibliography{draft}

\end{document}